\documentclass[aps, prb, twocolumn, floatfix]{revtex4}

\usepackage{graphicx}
\usepackage{multirow}

\begin{document}

\title{Chemical functionalization on planar polysilane and graphane }

\author{Ning Lu}
\author{Zhenyu Li}
\thanks{Corresponding author. E-mail: zyli@ustc.edu.cn}
\author{Jinlong Yang}

\affiliation{Hefei National Laboratory for Physical Sciences at
     Microscale,  University of Science and Technology of
     China, Hefei,  Anhui 230026, China}

\date{\today}

\begin{abstract}
Two dimensional materials are important for electronics
applications. A natural way for electronic structure engineering on
two dimensional systems is on-plane chemical functionalization.
Based on density functional theory, we study the electronic
structures of fluorine substituted planar polysilane and graphane.
We find that carbon and silicon present very different surface
chemistry. The indirect energy gap of planar polysilane turns to be
direct upon fluorine decoration, and the gap width is mainly
determined by fluorine coverage regardless of its distribution on
the surface. However, electronic structure of fluorine doped
graphane is very sensitive to the doping configuration, due to the
competition between antibonding states and nearly-free-electron
(NFE) states. With specific fluorine distribution pattern,
zero-dimensional and one-dimensional NFE states can be obtained. We
have also studied the chemical modification with -OH or -NH$_2$
group. Carbon seems to be too small to accommodate big functional
groups on its graphane skeleton with a high concentration.
\end{abstract}

\maketitle

\section{introduction}

Low-dimensional materials are expected to play a key role in
nanotechnology. For example, it is desirable to use two dimensional
materials to build electronics on flexible, lightweight, and cheap
substrates like plastics. \cite{Guo08} Silicon is the fundamental
semiconducting material in electronics. Due to its relatively large
size, it prefers to form four covalent $\sigma$ bonds in tetragonal
coordination (sp$^3$ hybridization) instead of forming strong planar
$\pi$ bonds through sp$^2$ hybridization. A stable two dimensional
silicon material is planar polysilane, \cite{Takeda94} which is
composed of a buckled hexagonal silicon framework and hydrogen atoms
on both sides to saturate dangling bonds. Layered polysilane has
been synthesized experimentally,\cite{Dahn93, si-exp} and due to
confinement effect, its electronic structure is different from bulk
silicon. \cite{n1, n2, n3}

After the successful experimental preparation of graphene,\cite{a1}
carbon based materials become a rising star for two dimensional
electronics. \cite{a2, a3} Compared to silicon, carbon have smaller
radius, and stable $\pi$ bonds can be formed between carbon atoms
within a plane. Therefore, the stablest two dimensional carbon
structure is a hexagonal lattice, namely graphene. Carriers in
graphene show massless Dirac Fermion behavior, and there is a
symmetry between electron and hole in graphene. Since graphene is a
semimetal, an energy gap must be opened before any 'graphenium'
microprocessor can be realized. A natural way for this purpose is to
convert the sp$^2$ bond in graphene to sp$^3$ bond via
hydrogenation. Such a carbon based sp$^3$ structure is called
graphane, the carbon analogue of planar polysilane, which was
theoretically predicted to have a finite energy gap. \cite{b2, b3}
Recently, graphane has been synthesized by hydrogen plasma treatment
\cite{bh1} or by electron-induced dissociation of hydrogen
silsesquioxane. \cite{Ryu08}

For various electronics applications, it is very desirable to have
the capability to fully control the energy gaps and electronic
structures of planar polysilane and graphane. On-plane chemical
functionalization is an important way towards this direction.
Hydroxy group \cite{n2, n3, Deak92} substitution on planar
polysilane has been studied. The product, siloxene, has a direct
energy gap instead of the indirect one in unsubstituted planar
polysilane. \cite{Deak92, n2} For two dimensional carbon based
materials, previous electronic structure engineering attempts by
chemical means mainly focuses on graphene nanoribbon edge
modification. \cite{edge1, edge2, edge3} However, on-plane
functionalization does not need to cut the graphene sheet, and it is
expected to be a more feasible way to realize electronics on
graphene (for example, by printing circuit on graphene). Although
graphene oxide (GO) represents an on-plane chemically modified
carbon material, it is mainly studied as an intermediate for
massively production of graphene, \cite{g2, g4, g6} and the detailed
structure of GO is still unclear. \cite{g1, g3, g5, g8, g9}

Despite its importance, a systematical study of on-plane chemical
functionalization for two dimensional silicon and carbon materials
is still unavailable. Most importantly, how the electronic structure
depends on the adsorption configuration of functional groups has not
been discussed, which is very important for practical realization of
electronic structure engineering in experiment. Another not
considered important issue is the nearly-free-electron (NFE) states.
NFE states are delocalized on the surface outside the atomic
centers. \cite{f1} When they are close to the band edge, NFE states
become a determining factor of the band gap width. \cite{f2, f3, f4}
Since NFE states have unique space distribution, we expect their
response to surface chemical functionalization to be different than
other states.

In this article, we systematically studied the on-plane chemical
modification on planar polysilane and graphane. Functionalization
leads to significant electronic structure modification.
Interestingly, electronic structure of fluorinated polysilane is
much less sensitive to the F adsorption configuration than graphane.
The remainder of the paper is organized as the following. In section
II, we describe the computational details. The main results are
discussed in section III. We first compare the electronic structure
of two dimensional polysilane and graphane, and the NFE state on
graphane is introduced. Then, we discuss different behaviors upon F
substitution for these two systems. Other functional groups are
finally discussed. Section IV concludes the paper.

\section{Computational Details}

To investigate the geometric structures and electronic states of
polysilane and graphane based systems, we performed first-principles
DFT calculations with the Vienna Ab initio Simulation Package
(VASP).\cite{n8, n9} Projector-augmented wave (PAW) method
\cite{n10, n11} was used for electron-ion interaction, and the
Perdew-Wang form \cite{n12} of the generalized gradient
approximation (GGA) was adopted for electronic exchange and
correlation.  A kinetic-energy cutoff of 400 eV was selected for
plane wave basis set. For pristine planar polysilane and graphane we
chose a 1$\times$1 unit cell, and a 2$\times$2 supercell was used to
consider chemical modification. The vacuum space between two
neighboring two-dimensional sheets is larger than 15 \AA, and dipole
correction is applied to compensate artificial dipole interaction
between neighoring supercells in the direction perpendicular to the
plane. A 13$\times$13$\times$1 $k$-point mesh was used in geometry
optimizations, and a 27$\times$27$\times$1 grid was used for static
electronic structure calculations. Geometry structures were relaxed
until the force on each atom was less than 0.01 eV/\AA\, and the the
convergence criteria for energy is $10^{-5}$ eV. A series of unit
cells were scanned for each system to obtained the lattice
parameters with the lowest energy. Charge populations were
calculated using Bader's atom in molecule (AIM) method based on
charge density topological analysis.\cite{d2}

\section{RESULTS AND DISCUSSION}

\subsection{Geometrical and electronic structures of planar polysilane and graphane}

\begin{figure}[tbhp]
\includegraphics[width=8cm]{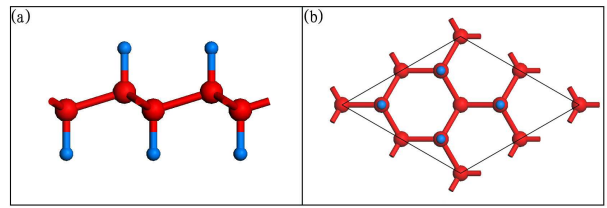}
\caption{Schematic (a) side view and (b) top view of planar
polysilane and graphane. Silicon or carbon is in red, and hydrogen
in blue. An 2$\times$2 unit cell is marked in the top view.}
\label{fig:geo}
\end{figure}

First, we compare the geometrical and electronic structures of
planar polysilane and graphane. Both of them are composed of a
buckled hexagonal plane of silicon or carbon, and hydrogen atoms on
both sides of the plane to saturate the sp$^3$ $\sigma$ bonds (Fig.
\ref{fig:geo}). The optimized Si-Si bond length is 2.36 \AA, similar
to its calculated bulk value (2.35 \AA). Si-H bond length is 1.50
\AA. The optimized C-C and C-H bond length in graphane is 1.54 and
1.11 \AA, respectively. The C-C bond length in graphane is almost
identical to its value in bulk diamond, and it is also in agreement
with the previous theoretical results on graphane. \cite{b2, b3}

\begin{figure}[tbhp]
\includegraphics[width=8cm]{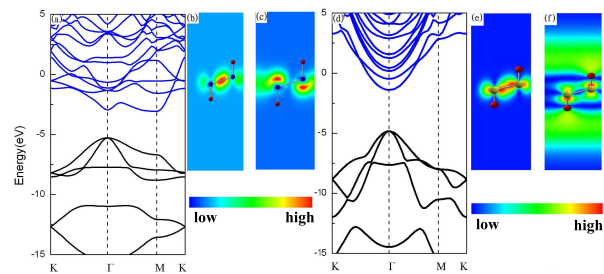}
\caption{(a) Band structure of planar polysilane, and corresponding
charge densities of (b) the VBM and (c) the conduction band edge at
the $\Gamma$ point. (d) Band structure of graphane, and
corresponding charge densities of (e) the VBM and (f) the CBM at the
$\Gamma$ point. In the band structures, occupied bands are in black,
and unoccupied bands are in blue.} \label{fig:Band-Si-C-H}
\end{figure}

Because its silicon or carbon atoms are sp$^3$ hybridized, both
planar polysilane and grphane have a significant energy gap (Figure
\ref{fig:Band-Si-C-H}). In agreement with Hirayama et al.,\cite{n3}
we obtained a quasi-direct band gap about 2.20 eV for two
dimensional polysilane, with the valance band maximum (VBM) at the
$\Gamma$ point and the conduction band minimum(CBM) at the M point.
The direct band gap at the $\Gamma$ point is just 0.12 eV broader.
Graphane has an about 3.43 eV direct gap at $\Gamma$, as also
suggested by Sofo et al. \cite{b2}

The VBM of polysilane is a Si-Si bonding state. Due to symmetry, it
is two-fold degenerated at the $\Gamma$ point. The conduction band
edge at $\Gamma$ point is the antibonding state between Si 3p$_z$
and H 1s orbitals, mainly distributed on Si. VBM of graphane have
similar bonding character with planar polysilane. However, CBM state
of graphane is notably different from that of ploysilane, where a
strongly delocalized feature exists above and blow the graphane
plane. Such a delocalized state is known as NFE state. \cite{f1}
When it is pushed to the band edge, NFE state becomes a determining
factor of the width of band gap. We note that the NFE-like character
of its CBM state has not been noticed in previous study on graphane.
\cite{b2}

Why CBM of graphane is an NFE state, while that of polysilane is
not? One reason is that the antibonding states of graphane are
higher in energy than those of planar polysilane. On the other hand,
NFE states of graphane are also strongly stabilized compared to
those of planar polysilane. In fact, at the $\Gamma$ point, the
lowest NFE states of polysilane and graphane locate at about 0.7 and
1.4 eV below the vaccum level, respectively. There are two possible
reasons for the stabilization of the NFE states in graphane. First,
compared to notably negatively charged hydrogen in polysilane,
hydrogen atoms in graphane are almost neutral. A negatively charged
surface is repulsive to electrons, and thus upshift NFE states.
Second, the radius of carbon is smaller than silicon, which leads to
a much smoother graphane surface than the polysilane surface.
Actually, the distance between two hydrogen atoms in graphane is
1.35 \AA\ shorter than that in polysilane. A smoother surface is
expected to stabilize NFE states.

\subsection{Fluorinated planar polysilane}

Fluorine has a much larger electronegativity than hydrogen.
Therefore, the substitution of hydrogen in planar polysilane by
fluorine may lead to a significant electronic structure
modification. We first consider the simplest case where all hydrogen
atoms are replaced by fluorine atoms (100\% F coverage). F
substitution does not change the Si-Si bond length, and the
optimized Si-F bond length is 1.63 \AA.

\begin{figure}[tbhp]
\includegraphics[width=8cm]{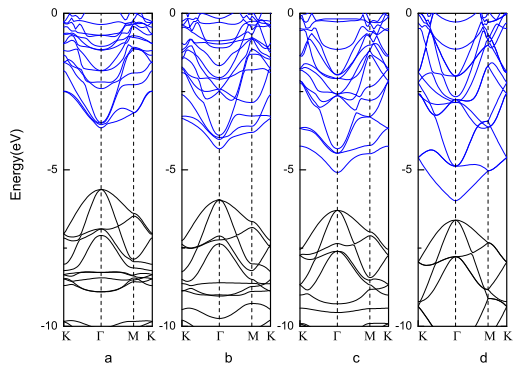}
\caption{Band structure of fluorinated planar polysilane for the
most stable configuration with a (a) 25\%, (b) 50\%, (c)75\%, and
(d) 100\% F coverage, respectively.} \label{fig:sif-4band}
\end{figure}

The band structure of F-substituted polysilane (Fig.
\ref{fig:sif-4band}d) is very different from the unsubstituted
system. The band gap (0.62 eV) becomes much smaller, and it is a
direct gap at the $\Gamma$ point. Therefore, F-substituted
polysilane will have a much higher luminous efficiency than bulk
silicon. CBM of F-substituted polysilane at the $\Gamma$ point is
mainly an antibonding state between Si and F molecular orbitals.
Similar to siloxene, \cite{n3} the indirect gap to direct gap
transition from unsubstituted to substituted polysilane may caused
by stronger mixing with F 2{\it s} state at $\Gamma$ point, which
pushes down the CBM at $\Gamma$ to lower than the local minimum at
the M point. VBM is still mainly a Si-Si bonding state. However,
compared to planar polysilane (Figure \ref{fig:Band-Si-C-H}c), there
is a small hybridization with the F 2{\it s} orbital.

\begin{figure}[tbhp]
\includegraphics[width=8cm]{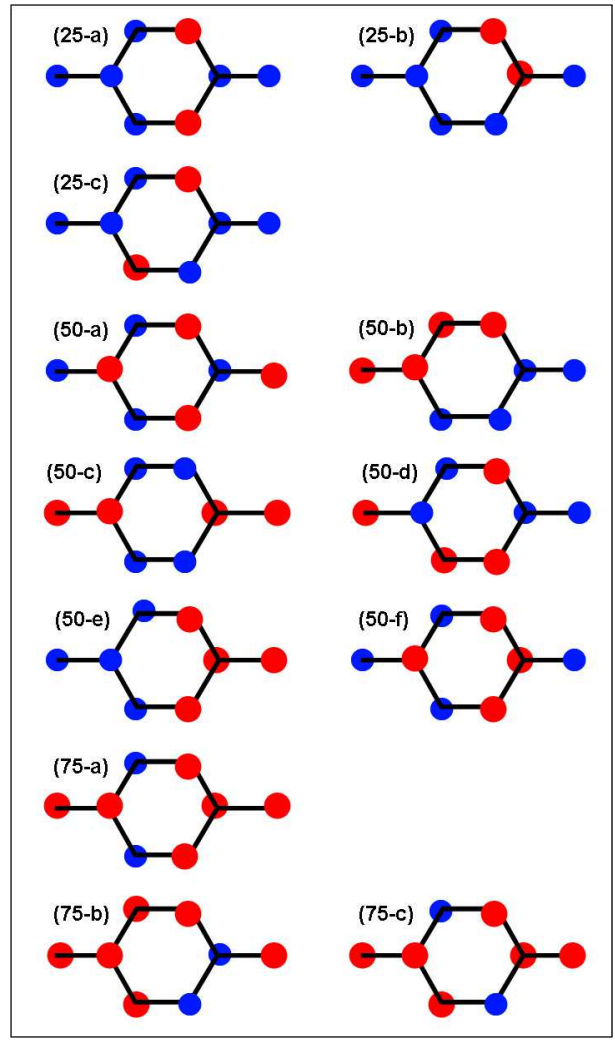}
\caption{Fluorine substitution configurations. Fluorine atoms are in
red, and hydrogen in blue. Black lines represent the silicon or
carbon skeleton.} \label{fig:sif+cf}
\end{figure}

It is also interesting to check the electronic structure evolution
with the F coverage. Using a 2$\times$2 supercell, we consider 25\%,
50\%, and, 75\% F substitutions, which corresponds to 2, 4, and 6
fluorine atoms in the supercell, respectively. For 25\% and 75\% F
coverages, we consider all possible F substitution configurations.
For 50\% F coverage, 6 configurations are considered. All calculated
configurations for different F concentrations are plotted in Fig.
\ref{fig:sif+cf}.

The energetically most stable configuration with different F
coverage is 25-c, 50-c, and 75-c, respectively. However, the energy
difference between different configurations for a certain F
concentration is very small, typically within 0.1 eV. At the same
time, with a certain F coverage, configuration difference has little
effect on the band structure. As shown in Fig. \ref{fig:sif-4band},
band gap decreases with the increase of the F coverage (Table
\ref{tb2:b}).

In all configurations, CBM is mainly of antibonding character
between Si-F or Si-H molecular orbitals. Since the Si-F antibonding
state is lower in energy than the Si-H antibonding state (refer to
the 100\% F coverage case and the unsubstituted case), more weight
of Si-F antibonding character leads to lower energy of the band.
Therefore, as the percent of F substitutions increases, the energy
of CBM significantly decreases.

VBM of fluorinated polysilane is a Si-Si bonding state with a small
hybridization with F orbitals. The energy of VBM also slightly
decrease with the F coverage, which can be understand by charge
transfer between Si and F. Since fluorine has a larger
electronegativity than silicon, an electron-transfer from Si to F is
expected. Charge population Bader analysis shows that the amount of
electrons transfer from Si to F increases with the increase of the F
concentration (Table \ref{tb2:b}). As VBM is mainly a Si-Si bonding
state, a positively charged Si skeleton will push down the VBM.
Because the downshift of CBM is more significant than VBM, we still
get smaller bang gap for strongly fluorinated polysilane.

\begin{table*}[bth]
\caption{Energy gap($E_g$), averaged charge population ($\rho$) on
Si or C, and dipole moment ($\mu$) of planar polysilane and graphane
with different degree of fluorine functionalization. $E_g$ is in eV,
$\rho$ in electron, and $\mu$ in e$\times$\AA\ per supercell.}
\label{tb2:b}
\begin{tabular}{ccccccccccccccccccccccccc}
 \hline\hline
       &  & 0 & 25-a & 25-b & 25-c & 50-a & 50-b & 50-c & 50-d & 50-e & 50-f & 75-a & 75-b & 75-c & 100   \\
 \hline
                Si-F  & $E_g$  & 2.20 & 1.94 & 1.86 & 1.96 & 1.67 & 1.37 & 1.62 & 1.52 & 1.49 & 1.57  & 1.13 & 1.06 & 1.21 & 0.62 \\
                      & $\rho$ & 3.46 & 3.37 & 3.38 & 3.38 & 3.32 & 3.32 & 3.32 & 3.32 & 3.32 & 3.32  & 3.26 & 3.27 & 3.26 & 3.19 \\
                      & $\mu$  & 0    & 0.30 & 0    & 0    & 0.60 & 0    & 0    & 0    & 0.30 & 0.30  & 0.30 & 0    & 0    & 0    \\
 \hline
               C-F   & $E_g$   & 3.43 & 3.43 & 4.94 & 4.86 & 2.60 & 4.56 &5.44 &5.16  &4.54  & 4.46  & 4.17 & 4.14 & 4.77 &3.10 \\
                     & $\rho$  & 3.98 & 3.87 & 3.88 & 3.88 & 3.71 & 3.73 & 3.74    &3.74  & 3.73 & 3.73  & 3.57 & 3.58 & 3.59 & 3.41 \\
                     & $\mu$   & 0    & 0.36 & 0    & 0    & 0.64 & 0    & 0       &0     &0.32  & 0.32  & 0.28 & 0    & 0    & 0 \\
 \hline\hline
\end{tabular}
\end{table*}

\subsection{Fluorine functionalization on graphane}

\begin{figure}[tbhp]
\includegraphics[width=8cm]{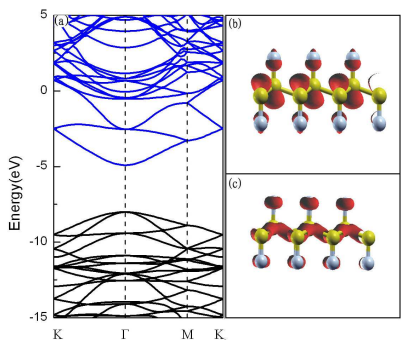}
\caption{(a) Band structure and charge densities of the (b) CBM and
(c) VBM at $\Gamma$ for fluorinated graphane.}
\label{fig:cf100-band+chg}
\end{figure}

If all hydrogen atoms in graphane are substituted by fluorine, the
C-C bond length will be slightly increased by 0.03 \AA\ to 1.57 \AA,
and the optimized C-F bond length is 1.38 \AA. As shown in Fig.
\ref{fig:cf100-band+chg}, F substitution decrease the band gap of
graphane to about 3.10 eV. CBM at the $\Gamma$ point is an
antibonding state between C and F orbitals, and VBM is a mix of C-C
bonding state and F 2{\it s} orbital. Compared to VBM of fluorinated
two dimensional polysilane, the hybridization with the F orbitals
here is stronger.

We also study different F coverage with all configurations listed in
Fig. \ref{fig:sif+cf}. For silicon, when the F coverage is fixed,
different substitution configurations lead to similar band
structures. Interestingly, the situation is totally different for F
partially substituted graphane, where the F ordering plays a key
role in the electronic structure of the doped systems.

\begin{figure}[tbhp]
\includegraphics[width=8cm]{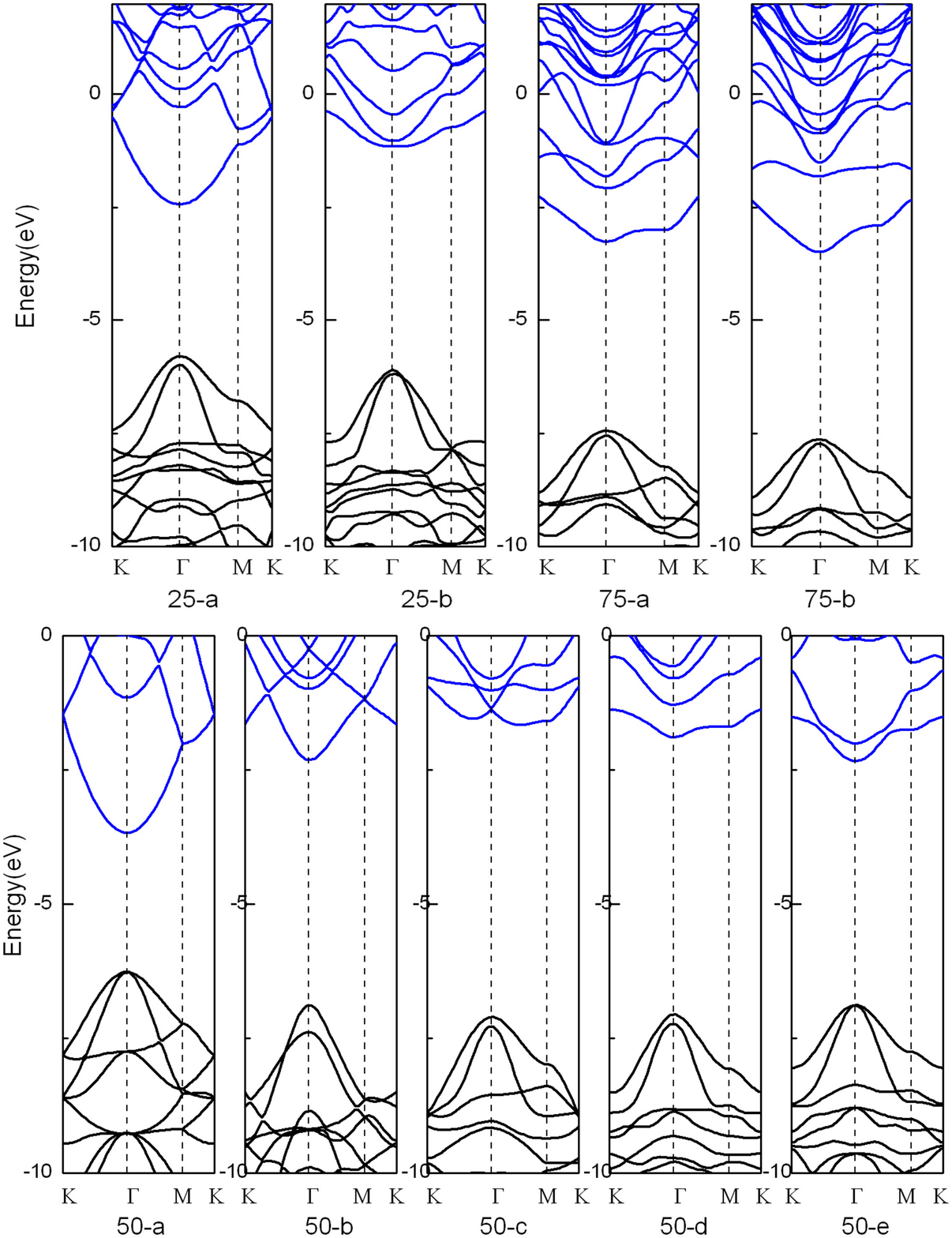}
\caption{Band structures of fluorinated graphane with different
configurations. } \label{fig:cf-coverageband}
\end{figure}

With the 25\% F coverage, we have three configurations. The 25-c
configuration has similar band structure with 25-b. Therefore, in
Fig. \ref{fig:cf-coverageband}, we only plot the band structures for
25-a and 25-b. There is an about 1.5 eV difference between the
energy gaps of 25-a and 25-b. VBM in both configurations is mainly a
C-C bonding state with a small mixing of F 2{\it p} orbital. Just
like F-substituted planar polysilane, its position slightly
decreases with the F coverage. The CBM of 25-a is significantly
lower than that of 25-b. CBM is mainly a NFE state similar to that
of unsubstituted graphane (Fig. \ref{fig:Band-Si-C-H}e). Since the
NFE state is delocalized apart from the system, its energy can be
easily affected by a shift of local vacuum level. A dipole layer on
surface can modify the local vacuum level. In Table \ref{tb2:b}, we
list the dipole moment for different configurations. Since, in 25-b
and 25-c, the two fluorine atoms is distributed on two sides of the
graphane plane, there is no net dipole moment for these two
configurations. However, in 25-a, the two fluorine atoms is on the
same side. Because fluorine atoms get electrons from the carbon
skeleton, there is a dipole moment perpendicular to the surface,
which lower the local vacuum level and thus also the NFE conduction
band.

For 50\% coverage, 50-f has similar electronic structure with 50-e,
and will not be discussed here. In the 50-a configuration, all F
atoms are on the same side, and the NFE-like CBM (distributed on the
H side) is lowered by the dipole layer generated by charge transfer
between C and F. As a result, the 50-a configuration has the
smallest energy gap in all 50\% F coverage configurations.

\begin{figure}[tbhp]
\includegraphics[width=8cm]{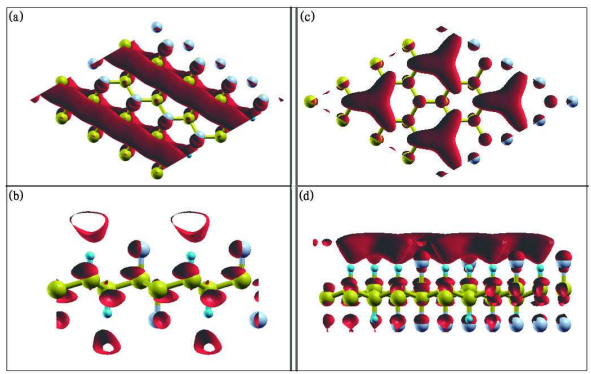}
\caption{(a) Top view and (b) side view of the lowest NFE state of
50-b. (c) Top view and (d) side view of the lowest NFE CBM state of
50-e.} \label{fig:chain+tri}
\end{figure}

CBM of the 50-b configuration is an antibonding state between C and
F molecular orbitals and the next two bands higher in energy at the
$\Gamma$ point are mainly NFE states (Fig. \ref{fig:chain+tri}a and
\ref{fig:chain+tri}b). Here, the lowest NFE state is higher than the
lowest antibonding state. This is because that the F doping destroy
the smooth H surface, which destabilize the NFE state. There is no
dipole moment in 50-b, and the dipole stabilization mechanism does
not help here. Interestingly, due to the chain configuration of H
atoms, the NFE state forms one-dimensional distribution. One
dimensional NFE state has potential application in electronics as an
ideal transport channel when it is close to the Fermi level.

Surprisingly, an indirect band gap is obtained for 50-c. At the
$\Gamma$ point, the lowest unoccupied band (LUB) is mainly a
ribbon-like NFE state, but also with a strong hybridization with
some C-H and C-F anti-bonding character. While the CBM at the
$k$-point between $\Gamma$ and M is mainly of antibonding character.
This fact indicates a strong interaction between the NFE state and
the antibonding state, which may be a reason for the indirect gap.
In 50-d configuration, we also have hydrogen chains formed, which
leads to two one-dimensional NFE bands (LUB+2 and LUB+3). However,
in this configuration, hydrogen chains on the two sides of the
graphane plane are not parallel.

Due to the dipole moment effect, CBM of 50-e is also a NFE state.
Because neighboring H atoms form triangles, NFE states in 50-e are
also of the shape of triangle (Fig. \ref{fig:chain+tri}c and
\ref{fig:chain+tri}d). Such zero-dimensional NFE states demonstrate
the flexibility to tune the distribution and energy of NFE states
over two dimensional surface by chemical functionalization. Due to
the smaller dipole moment compared to 50-a, the NFE state in 50-e is
higher, which leads to a larger gap.

The band structures of 75-a and 75-b are shown in Fig.
\ref{fig:cf-coverageband}, and the band structure of 75-c is similar
to 75-b. At the 75\% F coverage, due to the low coverage of
hydrogen, NFE state is not very stable, and CBM for all the three
configurations is antibonding states between C and F molecular
orbitals. Since the antibonding state is less affected by dipole
moment, the gap difference between different configurations at 75\%
F coverage is much less than those at 25\% or 50\% F coverages.

\subsection{Other functional groups}

Hydroxy (-OH) and amine (-NH$_2$) groups, both of which have
multiple atoms, are also considered. For simplicity, we only discuss
their 100\% coverage on both sides and 50\% coverage on one side
(full coverage on that side, the 50-a configuration).

OH or NH$_2$ functionalization does not change the lattice
parameters of two dimensional polysilane, which is consistent with
the relatively large Si-Si bond length and the week steric hindrance
between functional groups. However, for smaller carbon, due to the
repulsion between substituted radicals, the optimized C-C bond
length upon OH or NH$_2$ substitution is notably larger than that of
graphane. With 50\% and 100\% OH coverage, the C-C bond length is
1.57 and 1.62 \AA, respectively. For NH$_2$ functionalization, the
corresponding C-C bond length is 1.62 and 1.68 \AA, respectively.
The elongated C-C bond length indicates that it is difficult for
graphane to accommodate large functional groups at a relatively high
coverage, as also suggested by a previous GO structure
study.\cite{b3}

\begin{table}[bthp]
\caption{Energy gap($E_g$), averaged charge population ($\rho$) on
Si or C, and dipole moment ($\mu$) of planar polysilane and graphane
with different degree of OH and NH$_2$ functionalization. $E_g$ is
in eV, $\rho$ in electron, and $\mu$ in e$\times$\AA\ per supercell.
} \label{tb6:c}
\begin{tabular}{cccccccccc}
 \hline\hline
 & coverage  & 50\% & 100\% &  && coverage & 50\% & 100\%   \\
 \cline{1-4} \cline{6-9}
                Si-OH   & E$_g$  & 1.41 & 0.25       && C-OH     & E$_g$ & 1.85 & 2.21\\
                        & $\rho$ & 3.33 & 3.23         &         && $\rho$ & 3.77 & 3.49\\
                        & $\mu$ & 0.08 & 0             &         && $\mu$  & 0.25 & 0\\
 \cline{1-4} \cline{6-9}
                Si-NH$_2$      & E$_g$  & 1.33 & 0.37 && C-NH$_2$ & E$_g$ & 1.79 & 1.97\\
                               & $\rho$ & 3.37 & 3.28  &         && $\rho$ & 3.85 & 3.69\\
                               & $\mu$  & 0.30 & 0     &         && $\mu$ & 0.10 & 0\\

 \hline\hline
\end{tabular}
\end{table}

\begin{figure}[tbhp]
\includegraphics[width=8cm]{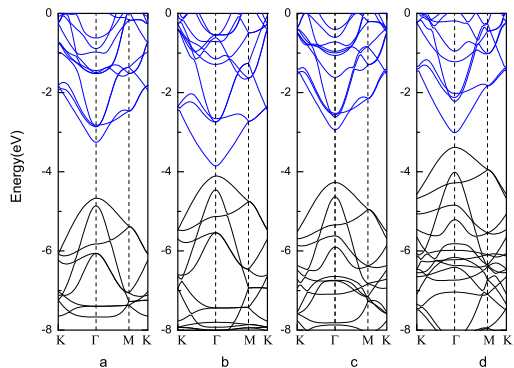}
\caption{Band structure of (a) 50\% and (b) 100\% OH covered or (c)
50\% and (d) 100\% NH$_2$ covered planar polysilane.}
\label{fig:sioh+sinh2}
\end{figure}

As shown in Fig. \ref{fig:sioh+sinh2}, the energy gap is also
decrease with the increase of the functional-group coverage for both
OH and NH$_2$ functionalized polysilane (Table \ref{tb6:c}).
However, contrast with the F substitution case, VBM rise with the
increase of the functional-group coverage. This may due to stronger
mixing between Si-Si bonding states and N or O 2$p$ orbitals, since
they are higher than F 2$p$ orbital and more close to the Si $3p$
orbital. The strong hybridization between Si and O or N orbitals is
consistent with the split VBM, which is degenerated in fluorinated
polysilane. CBM at the $\Gamma$ point is mainly of the antibonding
character between Si-O or Si-N molecular orbitals.

For both OH and NH$_2$ doped graphane, we obtain a smaller gap for
the 50\% coverage compared to the 100\% case. This is because the
NFE state on the hydrogen atom in the 50\% coverage case is lowered
by the dipole layer developed in the system. In all four cases, VBM
is a C-C bonding state mixed with O or N 2{\it p } orbitals. The
mixing is very strong, which leads to an even larger splitting of
VBM at the $\Gamma$ point compared to OH or NH$_2$ substituted
planar polysilane. The position of VBM is almost not affected by the
functional group coverage, which indicate that the N and O 2{\it p }
orbitals are close in energy to the C-C bonding state.

\begin{figure}[tbhp]
\includegraphics[width=8cm]{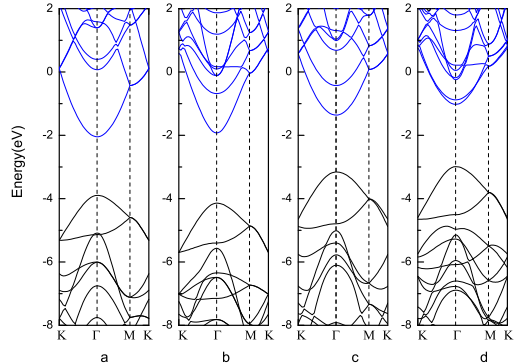}
\caption{Band structure of (a) 50\% and (b) 100\% OH covered or (c)
50\% and (d) 100\% NH$_2$ covered graphane.} \label{fig:coh+cnh2}
\end{figure}

\begin{figure}[tbhp]
\includegraphics[width=2.5cm]{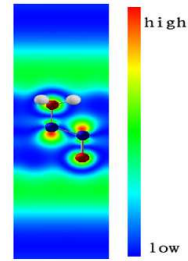}
\caption{CBM density in 100\% NH$_2$ covered graphane.}
\label{fig:cnh2-cbm}
\end{figure}

CBM of 100\% OH covered graphane is also the antibonding states of C
and O molecular orbitals. But for 100\% NH$_2$ covered graphane, CBM
has a significant NFE-like character due to the hydrogen atoms in
the radicals (Fig. \ref{fig:cnh2-cbm}). Such kind of NFE states
provide more flexibility to use NFE state to tune the electronic and
optical properties of two-dimensional materials.

\section{conclusions}

In summary, we have calculated the on-plane chemical modification of
two dimensional polysilane and graphane. Chemical functionalization
can change the indirect gap in planar polysilane to a direct gap,
which will strongly modify the optical properties of the system. For
fluorinated polysilane, the gap decreases with the F coverage,
independent on the F adsorption configurations. However, the
electronic structure of fluorinated graphane is sensitively
dependent on the F doping geometry. This is because the NFE states
of graphane is close to the conduction band edge. Energy of NFE
states is more sensitive to the local vacuum potential and the
doping structure. Functionalization by OH and NH$_2$ groups is also
discussed.

\begin{acknowledgements}
This work is partially supported by NSFC (20803071, 50721091,
20533030, 50731160010), by MOE (NCET-08-0521), by FANEDD (2007B23),
by the National Key Basic Research Program (2006CB922004), by the
USTC-HP HPC project, by the SCCAS and Shanghai Supercomputer Center.
\end{acknowledgements}

\end{document}